\documentstyle[12pt,epsfig,axodraw]{article}

\advance\textheight by \topskip
\newcommand{\be}{\begin{equation}}
\newcommand{\ee}{\end{equation}}
\newcommand{\br}{\begin{eqnarray}}
\newcommand{\er}{\end{eqnarray}}
\newcommand{\ba}{\begin{array}}
\newcommand{\ea}{\end{array}}
\newcommand{\bi}{\begin{itemize}}
\newcommand{\ei}{\end{itemize}}
\newcommand{\bn}{\begin{enumerate}}
\newcommand{\en}{\end{enumerate}}
\newcommand{\bc}{\begin{center}}
\newcommand{\ec}{\end{center}}

\newcommand{\Dir}{\kern -6.4pt\Big{/}}
\newcommand{\Dirin}{\kern -10.4pt\Big{/}\kern 4.4pt}
\newcommand{\DDir}{\kern -8.0pt\Big{/}}
\newcommand{\DGir}{\kern -6.0pt\Big{/}}

\def\slashchar#1{\setbox0=\hbox{$#1$}           
     \dimen0=\wd0                                 
     \setbox1=\hbox{/} \dimen1=\wd1               
     \ifdim\dimen0>\dimen1                        
        \rlap{\hbox to \dimen0{\hfil/\hfil}}      
        #1                                        
     \else                                        
        \rlap{\hbox to \dimen1{\hfil$#1$\hfil}}   
        /                                         
     \fi}                                         %

\def\be{\begin{equation}}
\def\ee{\end{equation}}
\def\bea{\begin{eqnarray}}
\def\eea{\end{eqnarray}}
\def\lsim{\:\raisebox{-0.5ex}{$\stackrel{\textstyle<}{\sim}$}\:}
\def\gsim{\:\raisebox{-0.5ex}{$\stackrel{\textstyle>}{\sim}$}\:}

\def\pl #1 #2 #3 {{Phys.~Lett.} {#1} (#2) #3}
\def\np #1 #2 #3 {{Nucl.~Phys.} {#1} (#2) #3}
\def\zp #1 #2 #3 {{Z.~Phys.} {#1} (#2) #3}
\def\jp #1 #2 #3 {{J.~Phys.} {#1} (#2) #3}
\def\pr #1 #2 #3 {{Phys.~Rev.} {#1} (#2) #3}
\def\prep #1 #2 #3 {{Phys.~Rep.} {#1} (#2) #3}
\def\prl #1 #2 #3 {{Phys.~Rev.~Lett.} {#1} (#2) #3}
\def\mpl #1 #2 #3 {{Mod.~Phys.~Lett.} {#1} (#2) #3}
\def\rmp #1 #2 #3 {{Rev. Mod. Phys.} {#1} (#2) #3}
\def\cpc #1 #2 #3 {{Comp. Phys. Commun.} {#1} (#2) #3}
\def\sjnp #1 #2 #3 {{Sov. J. Nucl. Phys.} {#1} (#2) #3}

\def\slash{/\kern -5pt}

\def\ims #1 {\ensuremath{M^2_{[#1]}}}

\def\ar{\to}

\def\s22w{s_{2W}^2}

\begin{document}
\tolerance=100000
\thispagestyle{empty}
\setcounter{page}{0}
 \begin{flushright}
{\large RAL-TR-1999-029}\\
{\large DTP/99/54}\\
{\large June 1999}\\
{\large Revised August 1999}\\
\end{flushright}

\vspace*{\fill}

\begin{center}
{\Large \bf Detecting Heavy Charged Higgs Bosons \\[0.5cm]
at the LHC with Four $b-$Quark Tags}\\[2.cm]

{\Large D.J. Miller$^1$, S. Moretti$^1$, D.P. Roy$^{2,3,4}$,
W.J. Stirling$^2$} \\[10mm] 
{\it $^1$Rutherford Appleton Laboratory,
Chilton, Didcot, Oxon OX11 0QX, UK} \\[5mm]
{\it $^2$Departments of Physics and Mathematical Sciences, 
University of Durham,  \\
 South Road, Durham DH1 3LE, UK} \\[5mm]
{\it $^3$Tata Institute of Fundamental Research, Mumbai - 400 005,
India}\\[5mm]
{\it $^4$LAPTH, B.P. 110, F-74941 Annecy-le-Vieux Cedex, France
}  \\[3mm] 
\end{center}

\vspace*{\fill}

\begin{abstract}{\small\noindent
We investigate the signature of a heavy charged Higgs boson
of the Minimal Supersymmetric Standard Model in the lepton plus multi-jet 
channel at the Large Hadron Collider with four $b$-tags. The signal
is the gluon-gluon fusion process $gg\to t\bar b H^-$, followed by  
the $H^-\to \bar t b$ decay, while the main background is 
from $gg\to t\bar t b\bar b$. We find that the two can be separated
effectively by kinematic cuts and mass reconstruction, but the 
signal size is not very large  in the end. Nonetheless, with a 
good $b$-tagging efficiency, $\epsilon_b\sim50\%$, this channel can 
provide a viable signature over
a limited but interesting range of the parameter space.
}
\end{abstract}

\vspace*{\fill}
\newpage
\setcounter{page}{1}

\section{Introduction}
\label{sec_intro}

The detection at the Large Hadron Collider (LHC) of
charged Higgs bosons would represent an
unequivocal signal of physics beyond the Standard Model (SM).
While the SM predicts only a neutral Higgs boson $\phi$, any two Higgs doublet
extension of it predicts a pair of charged Higgs bosons $H^\pm$
along with three neutral ones: the $CP-$even  $H$ and $h$ and the
$CP-$odd  $A$ \cite{one}.  This is true in particular for the
Minimal Supersymmetric extension of the Standard Model (MSSM).  While
the SM Higgs boson may be hard to distinguish from one of the neutral
Higgs bosons of the MSSM, the charged Higgs boson carries the unambiguous
hallmark of the Supersymmetric
(SUSY) Higgs sector.  Moreover, in contrast to the case of 
super-particles, whose signature depends sensitively upon 
assumptions regarding the $R-$parity status and the nature of the SUSY 
breaking, the signature of
$H^\pm$ bosons is fairly model independent.  Therefore, there has been
considerable interest in looking for SUSY signals via the associated Higgs
sector and in particular the charged Higgs bosons.  Furthermore,
the masses and couplings of all the MSSM Higgs bosons are
given at tree level in terms of only two parameters: i.e., 
one of the Higgs masses (e.g., 
$M_A$) and the ratio of the vacuum expectation values of the 
two Higgs doublets ($\tan\beta$).  Thus the experimental determination of the
$H^\pm$ mass $M_{H^\pm}$ would go a long way to
 quantitatively determining the MSSM 
Higgs sector. 

In the MSSM one has a lower mass limit at tree level, $M_{H^\pm} >
M_{W^\pm}$, which is not significantly modified by radiative corrections.
There is also a comparable experimental limit  from direct LEP2
searches for the charged Higgs boson \cite{two}.  In fact,
using the MSSM mass relations
one can get a strong indirect bound on $M_{H^\pm}$ in the low
$\tan\beta$ region from the LEP2 limit on $M_{h}$,
the mass of the lightest Higgs boson of the MSSM \cite{three}.  As shown
in \cite{four}, however, this constraint can be evaded in modest extensions 
of the MSSM
involving an additional singlet Higgs boson.  Therefore it does not preclude
direct searches for charged Higgs bosons in the low $\tan\beta$ region. 

The Tevatron and (especially) LHC 
hadron colliders offer the possibility to carry on the charged
Higgs boson search to higher masses 
because of their higher energy reach.  Here, the most prominent source
of $H^\pm$ production for $M_{H^\pm} < m_t$ (light charged Higgs boson)
is top quark decay, $t
\rightarrow bH^+$.  The corresponding branching ratio (BR) can easily 
be estimated from the relevant part of the MSSM Lagrangian
\be\label{vertextbH}
{\cal L} = {e \over \sqrt{2} M_{W^\pm} \sin\theta_W} H^+ 
(m_b \tan\beta ~\bar t b_R + m_t \cot\beta ~\bar t b_L),
\label{one}
\ee
written in the diagonal CKM matrix approximation.  It suggests a large
$\bar tbH^+$ Yukawa coupling and hence large branching fraction for the $t
\rightarrow bH^+$ decay in the regions of low as well as very high
$\tan\beta$, i.e.,
\be
\tan\beta \lsim 1 \ \ {\rm and} \ \ \tan\beta \gsim m_t/m_b.
\label{two}
\ee
Interestingly, these two regions are favoured by
$b-\tau$ unification for a related reason: i.e., assuming $m_b =
m_\tau$ at the GUT scale one needs a large $\bar t b H^+$ Yukawa
coupling contribution to the Renormalisation Group Equations
(RGEs) to control the rise of $m_b$ at 
low energy scales \cite{five}.  The dominant decay channels of
light charged Higgs bosons are
$H^+ \rightarrow c \bar s$ and $H^+ \rightarrow t^\ast \bar
b \rightarrow b\bar b W$
at $\tan\beta~\lsim~1$, while the $H^+ \rightarrow \tau^+ \nu_\tau$
decay dominates for $\tan\beta > 1$ \cite{six}.  It may be noted here that
there are already some modest limits on $M_{H^\pm}$ from the Tevatron
top quark data \cite{seven} in the two regions 
(\ref{two}).  The search can be extended over a wider region of $M_{H^\pm}$
and $\tan\beta$ at the forthcoming 
Tevatron upgrade (TeV-2), by exploiting the distinctive
$\tau$ polarization in $H^\pm$ decay \cite{eight,nine}.  Moreover the 
detection range can 
be enlarged to encompass practically the entire $M_{H^\pm} < m_t$ region at
the LHC \cite{nine,ten}.

We shall investigate here the prospect of charged Higgs searches at the LHC
in the opposite case, when $M_{H^\pm} > m_t$ (heavy charged Higgs boson).  
The dominant decay mode of such a particle
 is $H^\pm \rightarrow t\bar b$, which suffers from 
  a large QCD background.
Therefore, it is not surprising that detecting a charged Higgs
boson heavier than top has been generally regarded
as very hard\footnote{Recent new insights into the problem can be
found in Ref.~\cite{KOpol}, where the exploitation of top quark
polarisation effects was advocated. For an alternative approach,
using the much suppressed but somewhat cleaner $H^\pm\to \tau\nu$ decay 
channel, see Ref.~\cite{Kosuketau}. In the same spirit, the mode
$H^\pm\to W^\pm h$ is currently being investigated in Ref. \cite{Wh}.}.
The largest signal cross sections at the LHC are expected to come from the
associated production of $H^\pm$ with top (anti)quarks, followed by its 
dominant decay mode, i.e.,
\be
gb \rightarrow tH^- \rightarrow t\bar tb,
\label{three}
\ee
and
\be
gg \rightarrow t \bar b H^- \rightarrow t\bar t b\bar b.
\label{four}
\ee
After the decay of the top pair one expects 4 $b-$jets in (\ref{four}) 
and 3 in
(\ref{three}), where the accompanying $\bar b$ sea-quark does not take part in
the hard scattering and hence escapes detection by being produced close
to the beam direction.  The charged Higgs
signal from (\ref{three}) and (\ref{four}) were investigated 
in \cite{eleven} and \cite{twelve},
respectively, assuming triple $b$-tagging.  The dominant (QCD) background
processes in either case are
\be
gg \rightarrow t\bar t b\bar b,
\label{fivea}
\ee
\be
gg \rightarrow t\bar tgg~+~t\bar t q\bar q, 
\label{fiveb}
\ee
with one or more of the light parton jets in the
 latter process misidentified as
$b$.  Both the analyses found fairly viable $H^\pm$ signals in the two
above mentioned regions of $\tan\beta$, corresponding to a large $H^+ \bar t b$
Yukawa coupling.  Recently the associated production of charged Higgs
and $W^\pm$ bosons has been investigated in \cite{thirteen}.
Being a second order 
electroweak process, however, the size of the resulting signal is
smaller than those of reactions (\ref{three})--(\ref{four}). Furthermore,
it suffers from an overwhelming irreducible background induced by
top-antitop production and decay \cite{fourteen}.  
Therefore it does not seem to offer a useful $H^\pm$
signal at the LHC. Similarly, the production of charged
Higgs scalars in association with $b$ quarks \cite{newfourteen}
is burdened by a large combinatorial background as well as a formidable 
multi-jet QCD noise.

We shall present here an analysis of $H^\pm$ signals at the LHC produced
via the gluon-gluon fusion process (\ref{four})\footnote{For some early 
numerical 
studies of the on-shell $2\ar 3$ production, 
see Ref.~\cite{fifteen}.}, assuming all four $b-$quark jets to be
tagged.  Of course the size of this signal will be smaller than in
the $3b$-tagged channel. However, we have verified,
while computing the $gg$-initiated process, that,
by imposing a transverse momentum cut of 20 to 30 GeV on the spectator
$b-$jet in reaction (\ref{four}), the typical loss of signal is not
dramatic: e.g.,
about a factor of 2 to 3, 
for $M_{H^\pm}\approx300$ GeV. Besides, it is clear from the analyses in
Refs.~\cite{eleven,twelve}
 that the background (\ref{fiveb}) with {double} $b$-mis-tagging will be
very small. Indeed, we have explicitly confirmed this by recomputing the
processes (\ref{fivea})--(\ref{fiveb}) from scratch. The same conclusion
applies to the the case of single $b$-mis-tagging in events of the type: 
\begin{equation}\label{single}
gb\to t\bar t gb.
\end{equation} 
Therefore we only need to worry about the $t\bar t b\bar
b$ background (\ref{fivea}).  Moreover, we shall see below that the kinematics
of the pair of $b-$jets accompanying the $t\bar t$ pair is expected to be
rather different for the background, as compared to the signal (\ref{four}).
Consequently such a background can be suppressed effectively by using suitable
kinematic cuts on these $b-$jets.  As a result, we find a cleaner
(but smaller)  charged Higgs boson signal in the $4b$-tagged channel 
than in the
$3b$-ones  considered in \cite{eleven,twelve}.  In the
following Section 
we present the main steps in the calculation of the signal and background
cross sections. The event selection strategy and choice of
kinematic cuts are outlined in Section \ref{sec_selection}.  
The quantitative discussion of our analysis 
is presented in Section \ref{sec_results}.
Finally, we summarise the main results and 
present our conclusions in Section 
\ref{sec_summary}.

\section{Calculation of signal and background cross sections}
\label{sec_calcul}

The relatively large number of Feynman diagrams involved in the signal
process, $gg \rightarrow t\bar b H^-$ (\ref{four}), renders
the algebraic expressions for the corresponding `squared' matrix element
rather long and cumbersome. This drawback of the trace method
can be obviated by resorting to helicity amplitude techniques, which allow one
to write down the `complex' amplitudes in more compact form. In expressing
the helicity amplitudes we have made use of the formalism described, e.g., in 
Ref.~\cite{sixteen}, to which we refer the reader for technical
details. Here, we briefly 
outline the procedure and give all the formulae needed to implement our
calculations in a numerical program, as they have not appeared 
in the literature previously. 

First, one needs to introduce
some spinor functions, $Y$ and $Z$ ~\cite{seventeen,eighteen,ninteen}, 
given in Tabs.~\ref{tab_Y}--\ref{tab_Z},
which can be defined in terms of the following quantities 
($\epsilon^{0123} = 1$ is the Levi-Civita tensor):
\begin{eqnarray}\label{S}
S(+,p_1,p_2) &=& 2\, {{(p_1\cdot k_0)(p_2\cdot k_1)
 -(p_1\cdot k_1)(p_2\cdot k_0)
 +\mbox{i}\epsilon_{\mu\nu\rho\sigma}
  k^\mu_0k^\nu_1p^\rho_1p^\sigma_2}\over{\eta_1\eta_2}}, \\ \nonumber
S(-,p_1,p_2) &=& S(+,p_2,p_1)^*, \nonumber
\end{eqnarray}
\begin{equation}\label{mueta}
 \mu_i =\pm {m_i\over{\eta_i}}, \qquad\qquad
\eta_i =\sqrt{2(p_i\cdot k_0)}, 
\end{equation}
where $p_i$ and $m_i$ represent the four-momentum and mass
of the particle $i$ (for which $p_i^2=m_i^2$) 
and $k_0$  and $k_1$ are two arbitrary
four-vectors such that 
\begin{equation} 
k_0\cdot k_0=0, \quad\quad k_1\cdot k_1=-1, \quad\quad k_0\cdot k_1=0.
\end{equation}
In the first of the two expressions in eq.~(\ref{mueta}) the signs
$+$ and $-$ 
refer to  particles and to  antiparticles, respectively. 

\begin{table}[!t]
\begin{center}
\begin{tabular}{|c|c|}
\hline
$\lambda_1\lambda_2$ &
$Y(p_1,\lambda_1;p_2,\lambda_2;c_R,c_L)$ \\ \hline\hline
$++$  & $c_R\mu_1\eta_2+c_L\mu_2\eta_1$ \\ \hline
$+-$  & $c_LS(+,p_1,p_2)$ \\ \hline
\end{tabular}
\caption{The $Y$ functions for the two independent helicity combinations
in terms of the functions $S$, $\eta$ and $\mu$ defined in the text.
The remaining $Y$ functions can
 be obtained by flipping the  sign of the helicities
and exchanging $+$ with $-$ in the $S$ functions and $R$ with $L$ in the
chiral coefficients.}
\label{tab_Y}
\end{center}
\end{table}

\begin{table}[!t]
\begin{center}
\begin{tabular}{|c|c|}
\hline
$\lambda_1\lambda_2\lambda_3\lambda_4$ &
$Z(p_1,\lambda_1;p_2,\lambda_2;p_3,\lambda_3;p_4,
\lambda_4;c_R,c_L;c'_R,c'_L)$\\\hline \hline
$++++$ & $-2[S(+,p_3,p_1)S(-,p_4,p_2)c'_Rc_R
            -\mu_1\mu_2\eta_3\eta_4c'_Rc_L
            -\eta_1\eta_2\mu_3\mu_4c'_Lc_R]$  \\ \hline
$+++-$ & $-2\eta_2c_R[S(+,p_4,p_1)\mu_3c'_L-S(+,p_3,p_1)\mu_4c'_R]$ \\ \hline
$++-+$ & $-2\eta_1c_R[S(-,p_2,p_3)\mu_4c'_L-S(-,p_2,p_4)\mu_3c'_R]$ \\ \hline
$+-++$ & $-2\eta_4c'_R[S(+,p_3,p_1)\mu_2c_R-S(+,p_3,p_2)\mu_1c_L]$ \\ \hline
$++--$ & $-2[S(+,p_1,p_4)S(-,p_2,p_3)c'_Lc_R
            -\mu_1\mu_2\eta_3\eta_4c'_Lc_L
            -\eta_1\eta_2\mu_3\mu_4c'_Rc_R]$  \\ \hline
$+-+-$ & $0$ \\ \hline
$+--+$ & $-2[\mu_1\mu_4\eta_2\eta_3c'_Lc_L
+\mu_2\mu_3\eta_1\eta_4c'_Rc_R
-\mu_2\mu_4\eta_1\eta_3c'_Lc_R
-\mu_1\mu_3\eta_2\eta_4c'_Rc_L]$ \\ \hline
$+---$ & $-2\eta_3c'_L[S(+,p_2,p_4)\mu_1c_L-S(+,p_1,p_4)\mu_2c_R]$ \\ \hline
\end{tabular}
\caption{The $Z$ functions for all independent helicity combinations
in terms of the functions $S$, $\eta$ and $\mu$ defined in the text.
The remaining $Z$ functions can be obtained by
 flipping the sign of the helicities
and exchanging $+$ with $-$ in the $S$ functions and $R$ with $L$ in the
chiral coefficients.}
\label{tab_Z}
\end{center}
\end{table}

In Tabs.~\ref{tab_Y} and \ref{tab_Z} the (chiral) coefficients
$c_R$ and $c_L$ are those entering the fundamental fermion-fermion-boson
vertices, described through the expressions
\begin{equation}
\Gamma^{(')\mu}=\gamma^{\mu}\Gamma^{(')},
\end{equation}
and
\begin{equation}\label{vertex}
\Gamma^{(')}=c^{(')}_R P_R + c^{(')}_L P_L,
\end{equation}
with
\begin
{equation}P_R={{1+\gamma_5}\over{2}},\quad\quad\quad
P_L={{1-\gamma_5}\over{2}},
\end{equation}
the chiral projectors. 

If we make the following assignments for the momenta $p_i$ 
(with $i=1, ... 5$), which we assume incoming in the initial
state and outgoing in the final state, and helicities
$\lambda_i$ (with $i=1, ... 4$)
of the external particles in the $2\to3$ reaction 
\begin{equation}\label{proc}
g(p_1,\lambda_1)+     g(p_2,\lambda_2)
\rightarrow
t(p_3,\lambda_3)+\bar b(p_4,\lambda_4)+
H^- (p_5),
\end{equation}
so that $p_1^2\equiv p_2^2=0$, $p_3^2=m_t^2$ (with $\Gamma_3=\Gamma_t\ne0$), 
                               $p_4^2=m_b^2$ (with $\Gamma_4=\Gamma_b=0$)
                           and $p_5^2=M_{H^\pm}^2$,
then the Feynman amplitudes $T^{\{\lambda\}}_{i}$ can be written 
(apart from a phase, a factor $g_s^2 e$ and neglecting the colour matrices)
as (here and below, $\{\lambda\}$ refers cumulatively to 
the helicities of the external particles $\lambda_i$ with $i=1, ... 4$,
and $\sum_{\{\lambda\}}$ to the summation over all their possible 
combinations) 
%
%
\begin{eqnarray}\label{ggQQPhi1}
{T}^{\{\lambda\}}_{1}&=&
\begin{picture}(150,50)
\SetScale{1.0}
\SetWidth{1.2}
\SetOffset(0,-60)
\Gluon(45,75)(30,90){3}{3}
\Gluon(30,40)(45,55){3}{3}
\Text(25,95)[]{\small 1}
\Text(25,35)[]{\small 2}
\ArrowLine(60,40)(45,55)
\ArrowLine(45,55)(45,75)
\ArrowLine(45,75)(60,90)
\DashLine(55,85)(70,85){2}
\Text(77.5,85)[]{\small 5}
\Text(65,35)[]{\small 4}
\Text(65,95)[]{\small 3}
\Text(115,62.5)[]{=}
\end{picture} \\ \nonumber
&~& \;\;\;
\\ \nonumber
&~& \;\;\;
\\ \nonumber
&+&N_1 N_2 D_4(p_3+p_5)D_4(p_2-p_4)
                                 \sum_{i=1,2,4}\sum_{j=2,4}b_ib_j
                                 \sum_{\lambda=\pm}\sum_{\lambda'=\pm}
                                 \\ \nonumber
&&        Y(p_3,\lambda_3;p_i,\lambda; c_R^H,c_L^H)
\\ \nonumber
&\times& Z(p_i,\lambda;p_j,\lambda';
           p_1,\lambda_1;q_1,\lambda_1;
           c_R^g,c_L^g;
           1,1)
\\ \nonumber
&\times& Z(p_j,\lambda';p_4,-\lambda_4;
           p_2,\lambda_2;q_2,\lambda_2;
           c_R^g,c_L^g;
           1,1),
\nonumber
\end{eqnarray}
\begin{eqnarray}\label{ggQQPhi2}
{T}^{\{\lambda\}}_{2}&=&
\begin{picture}(150,50)
\SetScale{1.0}
\SetWidth{1.2}
\SetOffset(0,-60)
\Gluon(45,75)(30,90){3}{3}
\Gluon(30,40)(45,55){3}{3}
\Text(25,95)[]{\small 1}
\Text(25,35)[]{\small 2}
\ArrowLine(60,40)(45,55)
\ArrowLine(45,55)(45,75)
\ArrowLine(45,75)(60,90)
\DashLine(45,65)(60,65){2}
\Text(67.5,65)[]{\small 5}
\Text(65,35)[]{\small 4}
\Text(65,95)[]{\small 3}
\Text(115,62.5)[]{=}
\end{picture} \\ \nonumber
&~& \;\;\;
\\ \nonumber
&~& \;\;\;
\\ \nonumber
&-&N_1 N_2 D_3(p_3-p_1)D_4(p_2-p_4)
                                 \sum_{i=1,3}\sum_{j=2,4}b_ib_j
                                 \sum_{\lambda=\pm}\sum_{\lambda'=\pm}
                                 \\ \nonumber
&&       Z(p_3,\lambda_3;p_i,\lambda;
           p_1,\lambda_1;q_1,\lambda_1;
           c_R^g,c_L^g;
           1,1)
\\ \nonumber
&\times& Y(p_i,\lambda;p_j,\lambda'; c_R^H,c_L^H)
\\ \nonumber
&\times& Z(p_j,\lambda';p_4,-\lambda_4;
           p_2,\lambda_2;q_2,\lambda_2;
           c_R^g,c_L^g;
           1,1),
\nonumber
\end{eqnarray}
\begin{eqnarray}\label{ggQQPhi3}
{T}^{\{\lambda\}}_{3}&=&
\begin{picture}(150,50)
\SetScale{1.0}
\SetWidth{1.2}
\SetOffset(0,-60)
\Gluon(45,75)(30,90){3}{3}
\Gluon(30,40)(45,55){3}{3}
\Text(25,95)[]{\small 1}
\Text(25,35)[]{\small 2}
\ArrowLine(60,40)(45,55)
\ArrowLine(45,55)(45,75)
\ArrowLine(45,75)(60,90)
\DashLine(55,45)(70,45){2}
\Text(77.5,45)[]{\small 5}
\Text(65,35)[]{\small 4}
\Text(65,95)[]{\small 3}
\Text(115,62.5)[]{=}
\end{picture} \\ \nonumber
&~& \;\;\;
\\ \nonumber
&~& \;\;\;
\\ \nonumber
&+&N_1 N_2 D_3(p_3-p_1)D_3(p_4+p_5)
                                 \sum_{i=1,3}\sum_{j=1,2,3}b_ib_j
                                 \sum_{\lambda=\pm}\sum_{\lambda'=\pm}
                                 \\ \nonumber
&&       Z(p_3,\lambda_3;p_i,\lambda;
           p_1,\lambda_1;q_1,\lambda_1;
           c_R^g,c_L^g;
           1,1)
\\ \nonumber
&\times& Z(p_i,\lambda;p_j,\lambda';
           p_2,\lambda_2;q_2,\lambda_2;
           c_R^g,c_L^g;
           1,1)
\\ \nonumber
&\times& Y(p_j,\lambda';p_4,-\lambda_4; c_R^H,c_L^H),
\nonumber
\end{eqnarray}
\begin{eqnarray}\label{ggQQPhi4}
{T}^{\{\lambda\}}_{4}&=&
\begin{picture}(150,50)
\SetScale{1.0}
\SetWidth{1.2}
\SetOffset(0,-60)
\Gluon(45,75)(30,90){3}{3}
\Gluon(30,40)(45,55){3}{3}
\Text(25,95)[]{\small 2}
\Text(25,35)[]{\small 1}
\ArrowLine(60,40)(45,55)
\ArrowLine(45,55)(45,75)
\ArrowLine(45,75)(60,90)
\DashLine(55,85)(70,85){2}
\Text(77.5,85)[]{\small 5}
\Text(65,35)[]{\small 4}
\Text(65,95)[]{\small 3}
\Text(140,65)[]{= $~~~~~{T}^{\{\lambda\}}_{1}(1\leftrightarrow2)$,}
\end{picture} \\ \nonumber
&~& \;\;\;
\\ \nonumber
&~& \;\;\;
\nonumber
\end{eqnarray}
\begin{eqnarray}\label{ggQQPhi5}
{T}^{\{\lambda\}}_{5}&=&
\begin{picture}(150,50)
\SetScale{1.0}
\SetWidth{1.2}
\SetOffset(0,-60)
\Gluon(45,75)(30,90){3}{3}
\Gluon(30,40)(45,55){3}{3}
\Text(25,95)[]{\small 2}
\Text(25,35)[]{\small 1}
\ArrowLine(60,40)(45,55)
\ArrowLine(45,55)(45,75)
\ArrowLine(45,75)(60,90)
\DashLine(45,65)(60,65){2}
\Text(67.5,65)[]{\small 5}
\Text(65,35)[]{\small 4}
\Text(65,95)[]{\small 3}
\Text(140,65)[]{= $~~~~~{T}^{\{\lambda\}}_{2}(1\leftrightarrow2),$}
\end{picture} \\ \nonumber
&~& \;\;\;
\\ \nonumber
&~& \;\;\;
\nonumber   
\end{eqnarray}
\begin{eqnarray}\label{ggQQPhi6}
{T}^{\{\lambda\}}_{6}&=&
\begin{picture}(150,50)
\SetScale{1.0}
\SetWidth{1.2}
\SetOffset(0,-60)
\Gluon(45,75)(30,90){3}{3}
\Gluon(30,40)(45,55){3}{3}
\Text(25,95)[]{\small 2}
\Text(25,35)[]{\small 1}
\ArrowLine(60,40)(45,55)
\ArrowLine(45,55)(45,75)
\ArrowLine(45,75)(60,90)
\DashLine(55,45)(70,45){2}
\Text(77.5,45)[]{\small 5}
\Text(65,35)[]{\small 4}
\Text(65,95)[]{\small 3}
\Text(140,65)[]{= $~~~~~{T}^{\{\lambda\}}_{3}(1\leftrightarrow2),$}
\end{picture} \\ \nonumber
&~& \;\;\;
\\ \nonumber
&~& \;\;\;
\nonumber   
\end{eqnarray}
\begin{eqnarray}\label{ggQQPhi7}
{T}^{\{\lambda\}}_{7}&=&
\begin{picture}(150,50)
\SetScale{1.0}
\SetWidth{1.2}
\SetOffset(0,-62.5)
\Gluon(15,50)(30,65){3}{3}
\Gluon(30,65)(15,80){3}{3}
\Text(10,45)[]{\small 2}
\Text(10,85)[]{\small 1}
\Gluon(30,65)(60,65){3}{3}
\ArrowLine(75,50)(60,65)
\ArrowLine(60,65)(75,80)
\DashLine(70,75)(85,75){2}
\Text(92.5,75)[]{\small 5}
\Text(80,45)[]{\small 4}
\Text(80,85)[]{\small 3}
\Text(115,65)[]{=}
\end{picture} \\ \nonumber
&~& \;\;\;
\\ \nonumber
&~& \;\;\;
\\ \nonumber
&+&N_1 N_2 D_4(p_3+p_5)D(p_1+p_2)
                                 \sum_{i=1,2,4}b_i
                                 \sum_{\lambda=\pm}\sum_{\lambda'=\pm}
                                 \\ \nonumber
&&         Y(p_3,\lambda_3;p_i,\lambda; c_R^H,c_L^H)
\\ \nonumber
&\times&[  Y(p_i,\lambda;p_1;\lambda';1,1)
           Y(p_1,\lambda';p_4,-\lambda_4; c_R^g,c_L^g)
          \\ \nonumber
&&\times   Z(p_1,\lambda_1;q_1,\lambda_1;
             p_2,\lambda_2;q_2,\lambda_2;
             1,1;1,1)
           \\ \nonumber
&&+2~      Z(p_2,\lambda_2,q_2,\lambda_2;
             p_i,\lambda;p_4,-\lambda_4;
             1,1; 
             c_R^g,c_L^g)
          \\ \nonumber
&&\times   Y(p_1,\lambda_1;p_2;\lambda';1,1)
           Y(p_2,\lambda';q_1,\lambda_1;1,1)
          \\ \nonumber
&&-~{\mbox{same}}(1\leftrightarrow2)],
\nonumber 
\end{eqnarray}
\begin{eqnarray}\label{ggQQPhi8}
{T}^{\{\lambda\}}_{8}&=&
\begin{picture}(150,50)
\SetScale{1.0}
\SetWidth{1.2}
\SetOffset(0,-62.5)
\Gluon(15,50)(30,65){3}{3}
\Gluon(30,65)(15,80){3}{3}
\Text(10,45)[]{\small 2}
\Text(10,85)[]{\small 1}
\Gluon(30,65)(60,65){3}{3}
\ArrowLine(75,50)(60,65)
\ArrowLine(60,65)(75,80)
\DashLine(70,55)(85,55){2}
\Text(92.5,55)[]{\small 5}
\Text(80,45)[]{\small 4}
\Text(80,85)[]{\small 3}
\Text(115,65)[]{=}
\end{picture} \\ \nonumber
&~& \;\;\;
\\ \nonumber
&~& \;\;\;
\\ \nonumber
&-&N_1 N_2 D_3(p_4+p_5)D(p_1+p_2)
                                 \sum_{i=1,2,3}b_i
                                 \sum_{\lambda=\pm}\sum_{\lambda'=\pm}
                                 \\ \nonumber
&&      [  Y(p_3,\lambda_3;p_1;\lambda';1,1)
           Y(p_1,\lambda';p_i,\lambda; c_R^g,c_L^g)
          \\ \nonumber
&&\times   Z(p_1,\lambda_1;q_1,\lambda_1;
             p_2,\lambda_2;q_2,\lambda_2;
             1,1;1,1)
           \\ \nonumber
&&+2~      Z(p_2,\lambda_2,q_2,\lambda_2;
             p_3,\lambda_3;p_i,\lambda;
             1,1; 
             c_R^g,c_L^g)
          \\ \nonumber
&&\times   Y(p_1,\lambda_1;p_2;\lambda';1,1)
           Y(p_2,\lambda';q_1,\lambda_1;1,1)
          \\ \nonumber
&&-~{\mbox{same}}(1\leftrightarrow2)] 
\\ \nonumber
&\times&   Y(p_i,\lambda;p_4,-\lambda_4; c_R^H,c_L^H), 
\nonumber 
\end{eqnarray}
%
%
where we have introduced the coefficients 
$b_1=b_2=-b_3=-b_4=1$, to distinguish incoming and outgoing particles,
 and the propagators 
$D_{i}(p)=1/(p^2-m_{i}^2+{\mathrm{i}}m_i\gamma_i)$ 
(with $\gamma_i\equiv\Gamma_i$
if $p^2>0$ and $\gamma_i=0$ otherwise\footnote{That is, we include a finite
quark width only in resonant propagators.}) and $D(p)=1/p^2$.
In eqs.~(\ref{ggQQPhi1})--(\ref{ggQQPhi8}),
 $q_i$ (with $i=1,2$) is an arbitrary four-momentum
not parallel to $p_i$ (i.e., 
$q_i\ne\alpha p_i$ with $\alpha$ constant)
and $N_i=[4(q_i\cdot p_i)]^{-1/2}$; see Ref.~\cite{sixteen}
for more details.

The coefficients
$c_R$ and $c_L$ for the vertices
relevant to such a process are
\begin{equation}\label{chiralH}
(c_{R}^H,c_{L}^H)=-\frac{1}{\sqrt 2 M_{W^\pm}\sin\theta_W}
(m_b\tan\beta, m_t\cot\beta)
\end{equation}
for the charged Higgs, see eq.~(\ref{vertextbH}), and simply
\begin{equation}\label{chiralg}
(c_{R}^g,c_{L}^g)=(1,1)
\end{equation}
for the gluon to quarks couplings. 
As usual, $\theta_W$ represents the Weinberg angle. Here, 
the bottom and top quark masses entering the Yukawa interaction are
those defined at the propagator pole, 
i.e., $m_{b,t}\equiv \bar m_{b,t}(m_{b,t})$, where
$\bar m_{b,t}(Q)$ are the running masses at the (energy) scale $Q$ (see below).
 Also note that in
eqs.~(\ref{chiralH})--(\ref{chiralg}) we have factored out the overall
couplings $-\mbox{i}e$ and $-\mbox{i}g_s$ of the Lagrangian.

As for the colour structure of our process, one should notice that
in this case there are two basic combinations of colour matrices associated 
with the Feynman graphs (\ref{ggQQPhi1})--(\ref{ggQQPhi8}), that is
$(t^A t^B)_{i_3 i_4}$ and $(t^B t^A)_{i_3 i_4}$, where $A(i_3)$ and $B(i_4)$
identify the colours of the gluons(quarks) $1(3)$ and $2(4)$, respectively.
In fact, it should be recalled that the colour terms associated
with the triple-gluon vertices are nothing but the 
structure constants $f^{ABX}$ of the $SU(N_C)$ gauge group, for which
\begin{equation}\label{algebra}
[t^A,t^B]_{i_3i_4}\equiv (t^A t^B)_{i_3i_4}-(t^B t^A)_{i_3 i_4}={\mathrm{i}}
f^{ABX}t^X_{i_3i_4},
\end{equation}
$X$ being in our case the colour label of the virtual gluon. 
Therefore, one can conveniently group the original eight  
Feynman amplitudes as follows
\begin{eqnarray}\label{M_ggQQPhi}  
M^{\{\lambda\}}_{+}&=&\sum_{i=1}^3 T^{\{\lambda\}}_{i}
                     +\sum_{i=7}^8 T^{\{\lambda\}}_{i}
\\ \nonumber
M^{\{\lambda\}}_{-}&=&\sum_{i=4}^6 T^{\{\lambda\}}_{i}
                     -\sum_{i=7}^8 T^{\{\lambda\}}_{i}.
\nonumber
\end{eqnarray}
The standard form of the matrix element (ME) squared, summed/averaged over the
final/initial spins and colours,  is then
\begin{equation}
{|{\cal M}|^2}(gg \ar t\bar b H^-) =
  \frac{g_s^4 e^2}{256} 
\sum_{\{\lambda\}} \sum_{i,j=\pm}{M}^{\{\lambda\}}_i
M_j^{\{\lambda\}*} C_{ij},
\end{equation}
where $C_{ij}$ is a $2\times2$ colour matrix with elements
\begin{eqnarray}\label{colour}
C_{++}&\equiv&C_{--}=\frac{N_C}{4}(N_C^2-2+\frac{1}{N_C^2}) \\ \nonumber
C_{+-}&\equiv&C_{-+}=\frac{N_C}{4}(\frac{1}{N_C^2}-1),   \nonumber
\end{eqnarray}
$N_C=3$ being the number of colours in QCD. Finally, notice that the ME for the
charge conjugated process
\begin{equation}\label{ccproc}
g(p_1,\lambda_1)+     g(p_2,\lambda_2)
\rightarrow
b(p_3,\lambda_3)+\bar t(p_4,\lambda_4)+
H^+ (p_5),
\end{equation}
 now with $p_3^2=m_b^2$ and $p_4^2=m_t^2$ (and, correspondingly,
$\Gamma_3=0$ and $\Gamma_4=\Gamma_t$), can be obtained by the simple 
replacement 
$c_R^H\leftrightarrow c_L^H$.

The above formulae refer to the $2\ar 3$ process of on-shell $H^\pm-$ and 
$t-$production. In reality, both these particles eventually decay inside
the detectors,
so that one ought also to consider their decay signatures.
We have included these decays by convoluting the $2\ar 3$ (unpolarised) 
production ME with the $3-$ and $4-$body decay MEs of top and Higgs,
 respectively,
which are well known and can be found 
in the literature. In doing so, we introduce several 
simplifications\footnote{Note that
in the numerical simulations we allow the top quark and Higgs bosons 
to be `off-shell',
that is, $p_{3[4]}^2\ne m_t^2$ and $p_5^2\ne M_{H^\pm}^2$ in the above 
formulae for
process (\ref{proc})[(\ref{ccproc})].}.
Firstly, we neglect spin effects in the top decays.
Secondly, we do not consider interference effects between diagrams
 involving $H^+$
and $H^-$ production. Thirdly, Fermi-Dirac interferences 
due to indistinguishibility
of $b-$quarks (or, equivalently, of $\bar{b}-$antiquarks) 
in the final state are not 
included. However, while noticeably simplifying the numerical 
evaluation,  we have checked that
these approximations do not spoil the validity of our analysis. In fact, 
we have also produced the exact $2\ar 8$ ME for the signal
process, including all the above spin and interference effects,
by means of the {\tt HELAS} \cite{twenty} subroutines
and the MadGraph \cite{tim} package, and compared its 
yield to that
of the simplified implementation. We have always found a good agreement between
the two, with residual effects surviving only in differential spectra which are
of no concern in our analysis. Indeed, for the typical quantities 
we shall investigate
(transverse momenta, pseudorapidity, multi-jet invariant masses, etc.),  
the results generally coincide within numerical accuracy. 

The {\tt HELAS} libraries and MadGraph  
have also been used to generate the background process (5a), 
as any analytical
expression for the latter would be much too cumbersome. Indeed, 
thirty-six different
Feynman diagrams are involved (actually twice that if one considers also 
those induced
by the above mentioned Fermi-Dirac statistics), with ten external particles. 

Both signal and background MEs have been integrated by means of 
{\tt VEGAS} \cite{twentyone},
with a careful mapping of the phase space, to account for the various 
resonances.
In some cases, where the multi-dimensional integrations (21 in total, 
for the $2\ar8$
implementation) are more problematic, the {\tt VEGAS} results have been 
cross-checked
against those obtained by using {\tt RAMBO} \cite{twentytwo}
as well as the {\tt D01GCF} and {\tt D01GZF} {{\sc NAGLIB}} subroutines.  

In addition to the phase space integration, one also has to convolute in the 
$(x,Q^2)$-dependent Parton Distribution Functions (PDFs)
 for the the two incoming gluons. These have 
been evaluated at leading-order, by means of the packages 
MRS-LO(05A,09A,10A,01A,07A) \cite{twentythree}. The $Q^2$ used for the latter 
was the CM
energy (squared) at the parton level, i.e., $\hat s=x_1x_2s$, 
with $\sqrt s=14$~TeV taken 
as the CM energy for the LHC. The same choice has been adopted for
the scale of the strong coupling constant $\alpha_s$, evaluated again 
at lowest order,
with  $\Lambda_{\mathrm{QCD}}^{N_f=4}$ chosen in accordance 
with that of the PDF set used. 

In addition, notice that before the signal and background cross sections 
can be computed reliably, one must take into account the effects of 
higher-order QCD corrections to the tree-level
processes (see the discussion in Ref.~\cite{twelve}). The full next-to-leading 
order corrections are as yet unknown for the processes we consider.
For the case of the Higgs signal, the dominant effects
can however by mimicked by adopting
the pole masses in the Higgs-fermion coupling entering the production process,
see eq.~(\ref{chiralH}), rather than the running ones \cite{scott}.
As for the $t\bar t b\bar b$ background,  
we estimate their effect by including 
an overall $K$-factor of 1.5 throughout this study \cite{twelve}.

The numerical values of the SM parameters are
(pole masses are assumed):
$$m_\ell=m_{\nu_\ell}=m_u=m_d=m_s=m_c=0,
$$
$$\qquad m_b=4.25~{\mathrm{GeV}},
\qquad m_t=175~{\mathrm{GeV}},\ \ \ \ $$
$$M_Z=91.19~{\mathrm {GeV}},\quad\quad \Gamma_Z=2.5~{\mathrm {GeV}},$$
\begin{equation}\label{param}
M_{W^\pm}=80.23~{\mathrm {GeV}},\quad\quad \Gamma_W=2.08~{\mathrm {GeV}}.
\end{equation}
For the top width $\Gamma_t$, we have used the LO value calculated
within the MSSM (i.e., $\Gamma_t=1.55$ GeV for $M_{H^\pm} \gg m_t$).
Furthermore, we have adopted $M_{A}$ and $\tan\beta$ 
as independent input parameters defining the Higgs sector of the MSSM at LO.
The charged Higgs width $\Gamma_{H^\pm}$ has been computed by means of 
the program
{\tt HDECAY}, which indeed requires $M_A$, rather than $M_{H^\pm}$, as
mass input \cite{twentyfour}
(the masses of the super-partners of the ordinary matter
particles  were fixed well above 1~TeV,
so they  enter neither the top nor the charged Higgs decay chain
as real objects and render the virtual SUSY corrections 
to the $H^\pm tb$ vertex negligible \cite{sola}).
As this program uses running quark masses (in the modified Minimal 
Subtraction scheme)
in evaluating the decay width of the 
$H^\pm\ar tb$ channel we have, for consistency, used running values for
 all such quantities  in the $H^\pm tb$
couplings entering our decay MEs for the signal (but not in the propagators 
and in
the phase space, for which the pole masses $m_b\equiv\bar m_b(m_b)$ 
and $m_t\equiv\bar m_t(m_t)$ of 
eq.~(\ref{param})  have
been used).

Finally, notice that we stop our analyses at the parton level, without 
considering fragmentation and 
hadronisation effects. Thus, jets are identified with the
partons from which they originate and all cuts are applied directly to the
latter. In particular, when selecting $b-$jets, a vertex tagging is implied,
with finite efficiency $\epsilon_b$ per tag. As four $b-$jets 
will be required
to be tagged, the overall efficiency will be $\epsilon^4_b$, by which both
signal and background rates will eventually have to be multiplied. 
For simplicity,
 we assume no correlations between the four tags, nor do we 
include mis-identification of light-quark (including $c-$quark) jets
produced in the $W^\pm$ decay 
as $b-$jets. 
 
\section{Selection strategy}
\label{sec_selection}
 
In this Section we describe the kinematic procedure adopted 
to disentangle charged Higgs events (\ref{four}) from the background 
(\ref{fivea}) in the
$t\bar t b\bar b$ channel.  First, one of the top quarks is required to decay
leptonically ($t \rightarrow b\ell\nu$) to provide a hard lepton
($\ell=e,\mu$) 
trigger and avoid the QCD background, while the other decays
hadronically $(\bar t \rightarrow \bar b q q')$, with ($q,q'\ne b,t$).  
The resulting
branching fraction is $2 \times 2/9 \times 2/3$, with 
the factor of two accounting for the fact that each of the 
$W^\pm$'s can decay either leptonically or hadronically.  We assume that the
charge of the $b-$jet  
is not measured. Thus, the signature we are discussing is effectively 
\begin{equation}\label{signature}
4b~+~2~{\mathrm{jets}}~+~\ell^\pm~+~p^T_{\mathrm{miss}}, 
\end{equation}
where the 2 un-tagged jets and the $\ell + p^T_{\rm miss} (= p^T_\nu)$
arise from the $W^+W^-$ boson pair produced 
in the $t\bar t$ decay.

We note that all the decay products of the top quarks in the signal
are expected to be hard while one of the accompanying $b-$quarks 
(or both depending on the $M_{H^\pm}-m_t$ mass difference) could be soft. 
Therefore we
impose a relatively demanding transverse momentum cut on the two
softer $b-$jets,
\begin{equation}\label{pTj}
p^T_{b_1,b_2} > 20~{\mathrm{GeV}}.
\end{equation}
For simplicity, we do the same for the rest, i.e.,
\begin{equation}\label{pTl}
p^T_{\ell^\pm,\nu,j,b_3,b_4} > 20~{\mathrm{GeV}}.
\end{equation}
However, we will present some results also for the case of a 30~GeV cut
in transverse momentum (on both jets and leptons), since the latter
threshold is believed optimal in increasing the $b$-tagging efficiency
at high luminosity (this is needed in order to render the $4b$-signal of a 
charged Higgs boson statistically significant): see Ref.~\cite{notes}. 
Furthermore, we require the pseudorapidity of jets and leptons to
be  
\begin{equation}\label{eta}
|\eta_{b,j,\ell^\pm}| < 2.5,
\end{equation}
and  allow for their detection as separate objects by imposing
the following isolation criteria:
\begin{equation}\label{R}
\Delta R_{bb,bj,jj,b\ell^\pm,j\ell^\pm}>0.4,
\end{equation}
by means of  the variable
\begin{equation}\label{Rdef}
\Delta R_{ij}=\sqrt{(\Delta\eta_{ij})^2+(\Delta\phi_{ij})^2},
\end{equation}
defined in terms of relative differences in
pseudorapidity $\eta_{ij}$ and azimuth $\phi_{ij}$,
with $i\ne j=j,b,\ell^\pm$.

We simulate calorimeter resolution by a Gaussian smearing of $p^T$
(without shower spreading and with uniform pseudorapidity/azimuth 
segmentation),
with $(\sigma(p^T)/p^T)^2 = (0.6/\sqrt{p^T})^2 + (0.04)^2$ for all the
jets and 
$(\sigma(p^T)/p^T)^2 = (0.12/\sqrt{p^T})^2 + (0.01)^2$ for the leptons
\cite{eleven}.  The corresponding $p^T_{\rm miss}$ is evaluated from
the vector sum of the jet and lepton transverse momenta after
resolution smearing. 

To improve the signal/background ratio and to estimate the $H^\pm$
mass we follow a strategy similar to the one in Ref. \cite{eleven}. 
\begin{enumerate}
\item[{(a)}] The invariant mass of the two un-tagged jets is required
to be consistent with $M_{W^\pm}$,
\begin{equation}\label{thirtyfive}
|M_{jj} - M_{W^\pm} | \le  15 \ {\rm GeV}.
\end{equation}
\item[{(b)}] The neutrino momentum is reconstructed by equating
$p^T_\nu = p^T_{\rm miss}$ and fixing the longitudinal component
$p^L_\nu$ via the invariant mass constraint $M(\ell\nu) = M_{W^\pm}$.  The
latter gives two solutions.  If they are complex we discard the
imaginary parts and they coalesce; otherwise both the solutions are
retained. 
\item[{(c)}] The invariant mass formed by combining the un-tagged jet
pair with one of the four $b-$jets is required to match $m_t$,
\begin{equation}\label{thirtysix}
| M_{jjb} - m_t | \le 25 \ {\rm GeV}.
\end{equation}
If several $b-$jets satisfy this constraint, the one giving the best
agreement with $m_t$ is selected.
\item[{(d)}] The invariant mass formed by combining $\ell$ and $\nu$
with one of the 3 remaining $b-$jets is also required to match $m_t$
within $\pm 25 \ {\rm GeV}$.  If several $b-$jets satisfy this, the
one giving the best agreement with $m_t$ is selected along with the
corresponding value of $p^L_\nu$.
\item[{(e)}] The remaining pair of $b-$jets may be looked upon as the
$b\bar b$ pair accompanying the $t\bar t$ in the signal (\ref{four}) and
background (\ref{fivea}).  Note that one of these $b-$jets is expected to come
from the $H^\pm$ decay in the signal, while for the background they both
come from a gluon splitting.  Consequently, in the latter case one supposes 
the $b\bar b$ pair to have a smaller invariant mass.  
Furthermore, one may also expect the energy and the relative angle
of the two $b$'s to be rather 
different, between signal and background. 
We  compare the signal and background cross sections against $M_{bb}$,
$E_{b_1}$, $E_{b_2}$ (with the labels 1[2] referring to the more[less]
energetic of the two $b-$quarks) and $\cos\theta_{bb}$ and suppress
the latter by suitable cuts in one or more such quantities.
\item[{(f)}] Finally we combine each of the reconstructed $t-$quarks
with each of the remaining $b-$jets to obtain 4 entries for the
$bt$ invariant mass $M_{bt}$.  For each signal point, one of these
entries will correspond to the parent $H^\pm$ mass while the other
three will represent the combinatorial background.  We  plot the
signal and background cross sections against this quantity.  The
former will show the resonant peak at $M_{H^\pm}$ sitting on top of a
broad combinatorial background, while the latter will show only a
broad distribution in $M_{bt}$.  As we shall see below, the Breit-Wigner
peak itself can help to improve the signal/background ratio further as well
as to determine the $H^\pm$ mass.
\item[{(g)}] For $M_{H^\pm} \gg m_t$, one of the above mentioned
$b-$jets (i.e., the one coming from the $H^+ \rightarrow t\bar b$ decay)
would generally be much harder than the other.  In this case, one can expect to
reduce the
combinatorial background by combining each of the top quark pair with
the harder of the two accompanying 
$b-$jets.  This would give two values of invariant mass
$M_{bt}$ for each signal point, one of which corresponds to the
parent $H^\pm$ mass.  Therefore we shall also show the signal and
background cross sections against this quantity for $M_{H^\pm} \geq
300 \ {\rm GeV}$.
\end{enumerate}

\section{Results and discussion}
\label{sec_results}

\begin{figure}[p]
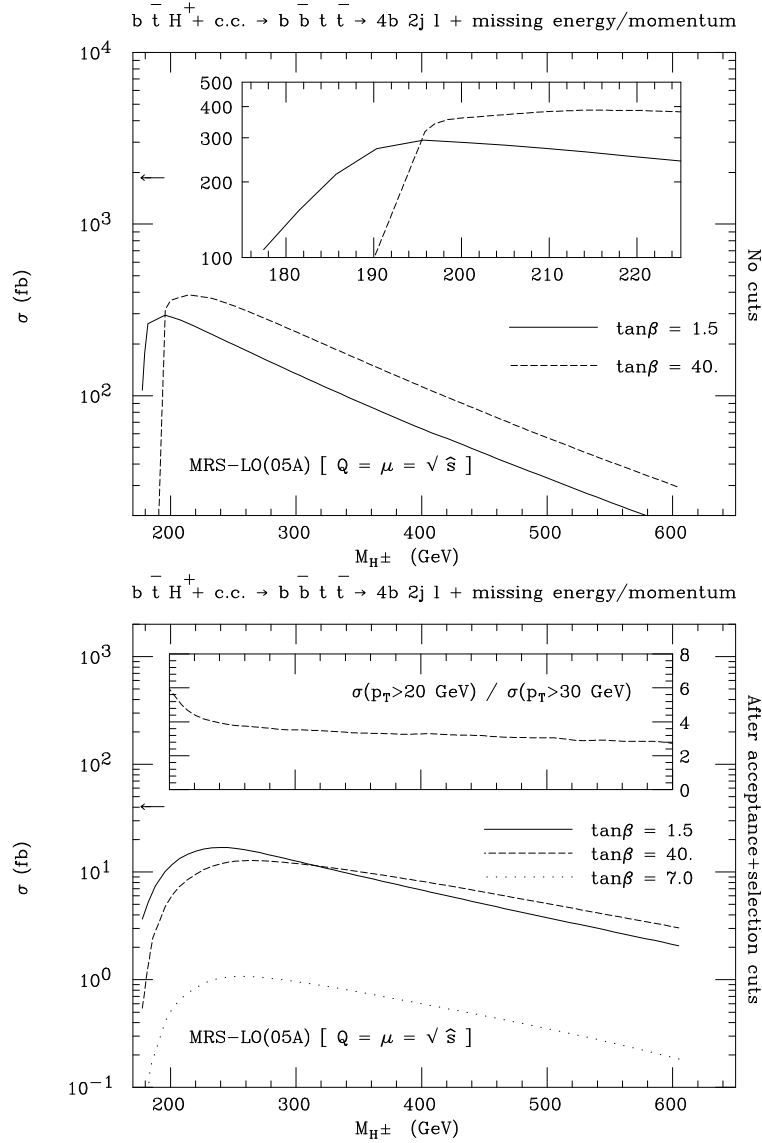

\begin{center}
{\epsfig{file=cross.ps,height=10cm,angle=90}}
\vskip0.001cm
{\epsfig{file=select.ps,height=10cm,angle=90}}
\caption{Production cross section for process (\ref{four}) (and its
charge conjugate) in the decay channel (\ref{signature})
as a function of $M_{H^\pm}$  in the heavy mass range,
for some discrete values of $\tan\beta$, for the case in which 
no (upper figure) and acceptance plus
selection (lower figure) cuts have been implemented. In the insert
 of the upper 
plot, we enlarge the rates in the vicinity of $M_{H\pm}=m_t$.
In the insert
 of the lower plot, we present the ratio between the above signal cross section
 for $\tan\beta=40$ and the corresponding one obtained by adopting a threshold
of 30 GeV in (\ref{pTj})--(\ref{pTl}).
The PDF set used was MRS-LO(05A)
with renormalisation and factorisation scales set equal to the partonic
CM energy. The arrow represents the size of the background
(\ref{fivea}) yielding the same signature (\ref{signature}).
No $b$-tagging efficiency is included.}
\label{fig:cross}
\end{center}
\end{figure}

Both signal (\ref{four}) and background (\ref{fivea}) cross sections
are finite over the entire phase space, provided the $b-$quark mass
is retained in the calculation. Thus, as a preliminary exercise,
we look at the total production rates of the above processes with
no cuts whatsoever, as they would appear 
in the decay channel (\ref{signature}) to an ideal detector. 
This is done in the top plot of Fig.~\ref{fig:cross}.
Here, the rates have been obtained by multiplying the $2\to3$ cross section
times the relevant BRs of top and Higgs decays, thus neglecting finite
width effects.
The signal rates depend on both $M_{H^\pm}$ and $\tan\beta$, and so 
they are plotted as a function of the former for two values
of the latter
in the favourable regions (\ref{two}). 
In contrast, the background rates are independent of
both and are indicated by the arrow next to the $y-$axis. The insert
in the top plot of 
Fig.~\ref{fig:cross} enlarges the region around the threshold region
$M_{H^\pm}\approx m_t$, where the BR($H^+\ar t\bar b$) increases rapidly.
A striking feature of the production cross sections is the apparently poor
signal-to-background ratio, over all the $M_{H^\pm}$ spectrum considered, 
irrespective of $\tan\beta$. Note however that no
dedicated treatment of the final state kinematics has yet been performed.
Indeed, the number of heavy charged Higgs events produced is sizable up
to around 800~GeV, where the total cross section at both $\tan\beta$ values is
still around several femtobarns. At its maximum, just after the opening of the
$H^+\ar t\bar b$ decay threshold, it can be larger by about two orders of
magnitude.

Fig.~\ref{fig:cross} has been produced by using the  
MRS-LO(05A)  PDF set. In Tab.~\ref{tab:PDF} we study the dependence
of both signal and background rates on the choice of the structure functions,
using the other four 
fits of the 1998 Martin-Roberts-Stirling-Thorne 
LO package\footnote{The additional
fits correspond to varying the large-$x$ gluon and $\alpha_s$
about their central preferred values.}. 
For reference, we have used the value $\tan\beta=30$, whereas
five different choices of Higgs masses in the heavy range
have been adopted.  Differences in  the
signal cross sections are found to be within the $\pm25\%$ range,  
with a somewhat smaller range
for the background. Furthermore,  changing the renormalisation ($Q$)
 and factorisation ($\mu$) scales from their common default value
$\sqrt{\hat s}$ to, e.g., the sum of the rest masses in the
$2\to3$ and $2\to4$ production process
(\ref{four}) and (\ref{fivea}), induces variations
 in the results of the same order as above. 
As a consequence, an overall error of, say, approximately 30--35\% 
should be taken as an estimate of the uncertainties related to the
PDFs and $\alpha_s$ throughout the paper.

As the next step of our analysis, we implement the acceptance cuts
(\ref{pTj})--(\ref{R}) and the selection cuts described through 
steps (a)--(d) in the previous Section. The total signal
and background  cross sections
 after such constraints have been enforced
can be found in the 
bottom plot of Fig.~\ref{fig:cross}. Note that the
kinematic procedure outlined above has been helpful in  increasing the 
signal-to-background ratio over all the Higgs mass spectrum
(compare to the curves above). 
The signal rates have been depleted too, mainly by the $p^T_{b_2}$
cut (\ref{pTj}), dropping to a few femtobarns for values of $\tan\beta$
at the upper and lower end of the parameter range and Higgs masses
below 700~GeV or so.  
For intermediate values of $\tan\beta$, e.g., around the
minimum of the production cross section occurring at 
$\tan\beta\approx7$, prospects are more gloomy. In fact,
the signal rates are always below 1~fb or so in this case,
for any $M_{H^\pm}$ value. Furthermore, as here
(and in the following as well) we have retained a finite value
for $\Gamma_H$, the reader may appreciate -- by comparing this plot to the
one above -- the suppressing(enhancing) effects 
of a larger(smaller) Higgs width at low(high) $M_{H^\pm}$ values
for $\tan\beta=40$, with respect  to the case $\tan\beta=1.5$
(the effects of a $\Gamma_t\ne0$ are the same in all cases). 
As intimated in the previous Section, we also have  considered
the case of a 30 GeV cut in all transverse momenta (including the
missing one). The insert
in the bottom plot of 
Fig.~\ref{fig:cross} presents the consequent suppression on the signal
rates for, e.g., 
$\tan\beta=40$. Far from the $M_{H^\pm}\approx m_t+m_b$ threshold,
the ratio between the two cross sections
stabilises at just below three (irrespectively of $\tan\beta$). 
For the background, the reduction
is slightly higher, a factor of four or so.

\begin{figure}[t!]
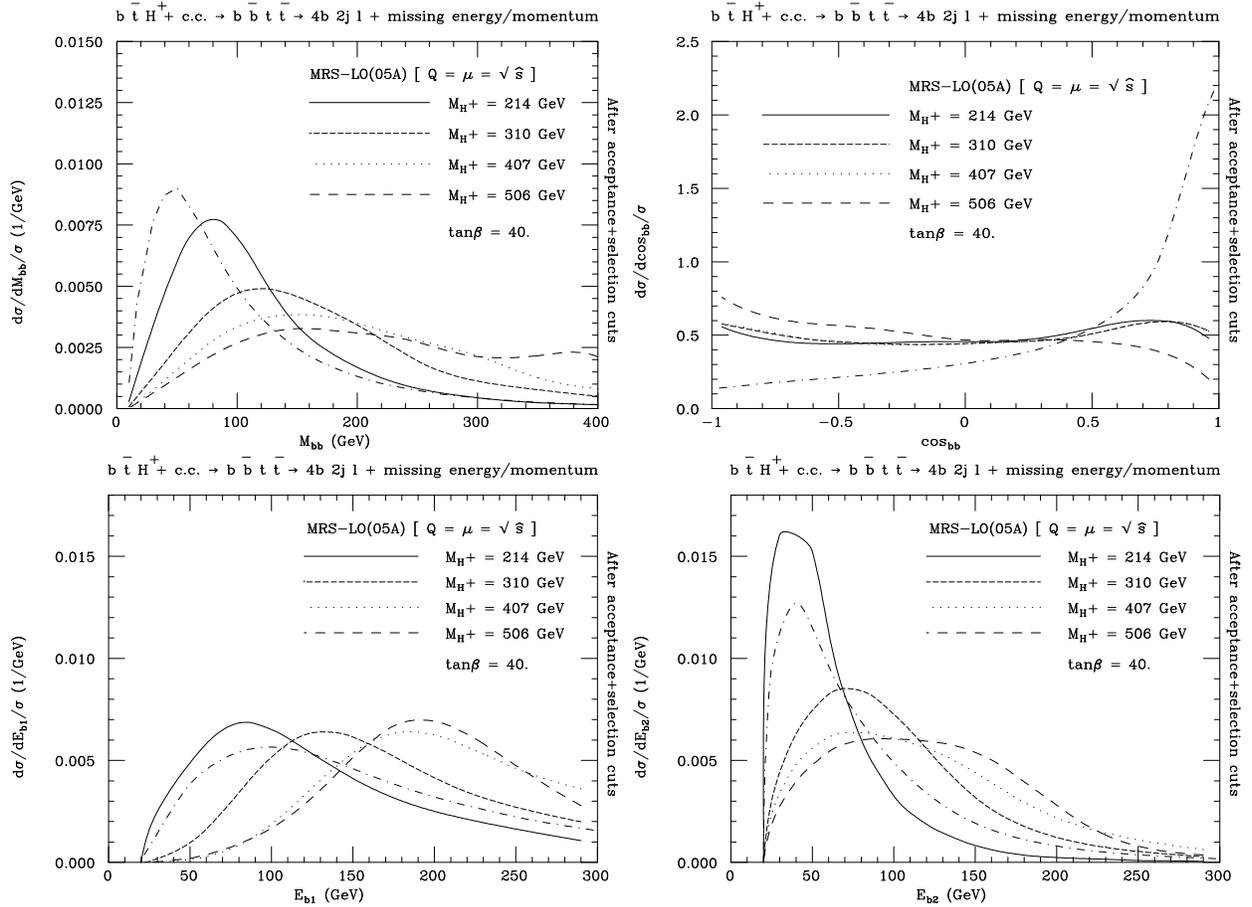

\begin{minipage}[b]{.495\linewidth}
\centering\epsfig{file=Mbb.ps,angle=90,height=6cm,width=\linewidth}
\end{minipage}\hfill\hfill
\begin{minipage}[b]{.495\linewidth}
\centering\epsfig{file=cosbb.ps,angle=90,height=6cm,width=\linewidth}
\end{minipage}\hfill\hfill
\begin{minipage}[b]{.495\linewidth}
\centering\epsfig{file=Eb1.ps,angle=90,height=6cm,width=\linewidth}
\end{minipage}\hfill\hfill
\begin{minipage}[b]{.495\linewidth}
\centering\epsfig{file=Eb2.ps,angle=90,height=6cm,width=\linewidth}
\end{minipage}
\caption{Differential distributions in invariant mass (top-left),
in cosine of the relative angle (top-right) and
in energy of the most (bottom-left) and least (bottom-right)
energetic of the two $b$ quarks accompanying the $t\bar t$ pair,
for process (\ref{four}) (and its
charge conjugate) in the decay channel (\ref{signature}),
for four selected values of $M_{H^\pm}$ in the heavy mass range,
with $\tan\beta=40$.
Acceptance and 
selection cuts have been implemented here. The PDF set used was MRS-LO(05A)
with renormalisation and factorisation scales set equal to the partonic
CM energy. The fifth (dot-dashed) curve represents the shape of the background
(\ref{fivea}) yielding the same signature (\ref{signature}). 
All distributions are normalised  to unity. 
No $b$-tagging efficiency is included.}
\label{fig:bb}
\end{figure}

Continuing with step (e) of our plan, we next investigate
the  mass, angular and energy behaviours of the two $b-$quarks
accompanying the $t\bar t$ pair, after the acceptance and selection cuts
have been implemented. 
The relevant plots can be found in Fig.~\ref{fig:bb}.
For reference, the
value chosen for $\tan\beta$ is 40, whereas for the charged
Higgs masses we have taken $M_{H^\pm}=214(310)[407]\{506\}$~GeV,
corresponding to $M_A=200(300)[400]\{500\}$~GeV. (Spectra look rather
similar if a $p^T_{\ell^\pm,\nu,{\mathrm{j}}},b>30$ GeV cut
is enforced instead.)
As already foreseen,
one can appreciate significant differences in all four quantities
considered, as long as $M_{H^\pm}$ is well above $m_t$. 
If $M_{H^\pm}$ is not much larger than 200~GeV, the  signal
and background distributions are similar except for the angular variable.
In contrast, if $M_{H^\pm}\ge 300$~GeV, one can see a greater
discriminatory power in each of these variables.

In order to enhance the signal-to-background ratio, especially
in the very heavy Higgs mass region, we therefore adopt the following
additional constraints on the  two $b-$jet system:
\begin{equation}\label{bbcuts}
M_{bb}>120~{\mathrm{GeV}},
\qquad
\qquad
\cos\theta_{bb}<0.75,
\qquad
\qquad
E_{b_1}>120~{\mathrm{GeV}}.
\end{equation}

\begin{figure}[p]
\begin{center}
\epsfig{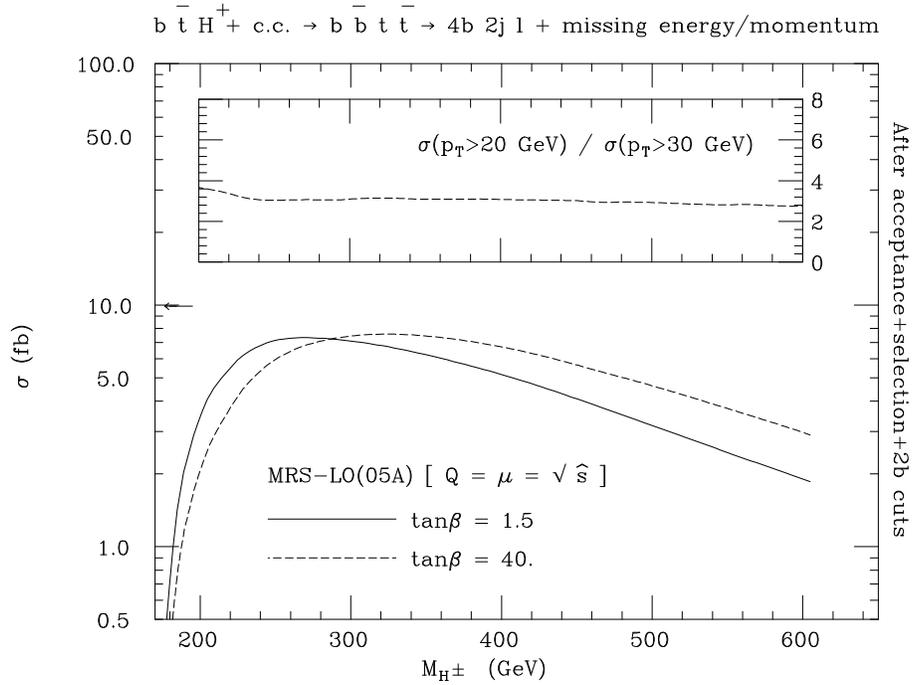}
\caption{Production cross section for process (\ref{four}) (and its
charge conjugate) in the decay channel (\ref{signature})
as a function of $M_{H^\pm}$ in the heavy mass range,
for two discrete values of $\tan\beta$.
Acceptance and selection cuts have been implemented here,
along with the additional cuts (\ref{bbcuts})
 on the $2b-$system accompanying the $t\bar t$ pair. 
In the insert, we present the ratio between the above signal cross section
 for $\tan\beta=40$ and the corresponding one obtained by adopting a threshold
of 30 GeV in (\ref{pTj})--(\ref{pTl}).
The PDF set used was MRS-LO(05A)
with renormalisation and factorisation scales set equal to the partonic
CM energy. The arrow represents the size of the background
(\ref{fivea}) yielding the same signature (\ref{signature}).
No $b$-tagging efficiency is included.}
\label{fig:final}
\end{center}
\end{figure}

The resulting signal cross sections are shown in Fig.~\ref{fig:final}
for $\tan\beta=1.5$ and 40. along with those of the background.
The signal-to-background ratio has increased, but the background
remains larger. However, notice the very little loss of signal
at very large Higgs masses. In fact, the Higgs rates remain above 1~fb for 
$M_{H^\pm}\lsim~800$ GeV. For a transverse momentum cut of 30 GeV throughout, 
the typical suppression on the signal (away from threshold) is again
about a factor of 3 (see the insert in the figure), now similar to the case 
of the background. 

\begin{figure}[p]
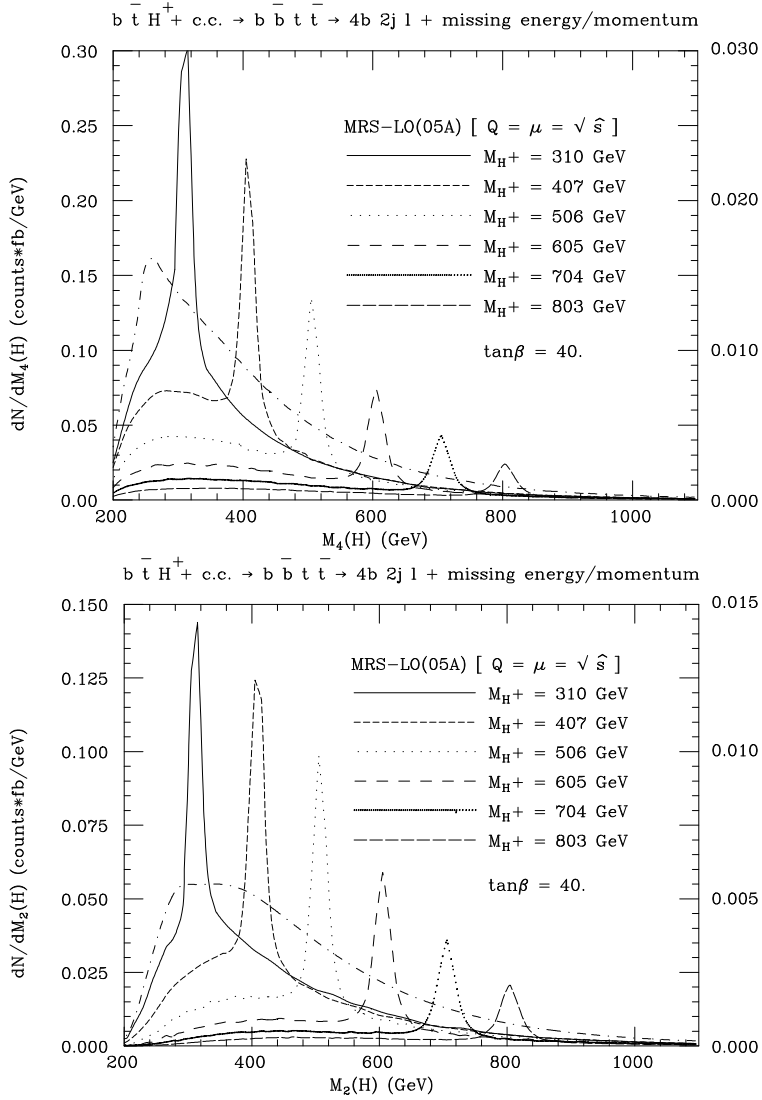

\begin{center}
{\epsfig{file=reso4.ps,height=10cm,angle=90}}
\vskip0.001cm
{\epsfig{file=reso2.ps,height=10cm,angle=90}}
\caption{Differential distributions in the reconstructed charged
Higgs mass for process (\ref{four}) (and its
charge conjugate) in the decay channel (\ref{signature}) for
six selected values of $M_{H^\pm}$ in the heavy mass range,
for $\tan\beta=40$. 
Acceptance and selection cuts have been implemented here,
along with the additional cuts (\ref{bbcuts})
 on the $2b-$system accompanying the $t\bar t$ pair. 
 The PDF set used was MRS-LO(05A)
with renormalisation and factorisation scales set equal to the partonic
CM energy. The seventh (dot-dashed) curve represents the shape of the 
background (\ref{fivea}) yielding the same signature (\ref{signature}). 
Normalisation is to the total cross sections times the number of
possible `$2~b\, +\, 2$~jet mass' combinations: four (top) and two (bottom).
The right-hand scale corresponds to a $b$-tagging efficiency factor of
$\epsilon_b^4=0.1$, i.e., $\epsilon_b=56\%$.}
\label{fig:reso}
\end{center}
\end{figure}

To enhance the relative rates further and estimate 
the charged Higgs boson mass, we reconstruct the $M_{bt}$ invariant mass
by combining each of the reconstructed top quarks with each of the
accompanying $b-$jets, as described in step (f), a quantity that
we label $M_4(H)$. The resulting spectra are presented
in the top plot of Fig.~\ref{fig:reso} for $\tan\beta=40$ and three selected
values of the $H^\pm$ mass along with the background. The signal
distribution clearly shows a resonance at $M_4(H)\approx M_{H^\pm}$
sitting over a combinatorial continuum, while the background spectrum is
broader and tends to concentrate at values of $M_4(H)$ below 300~GeV.
One can sharpen the resonances at the cost of reducing the size
of the signal by combining each of the reconstructed $t-$quarks
with the harder $b-$jet, as in step (g) (the corresponding invariant mass
is labeled as $M_2(H)$).
The distributions that we obtain in this way are given in the bottom part of
Fig.~\ref{fig:reso}.

One sees from the figure that the signal-to-background ratio is (much)
greater than one in the neighborhood of the resonant peaks. 
This is an advantage of the $4b-$tagged channel over the $3b-$tagged case
considered in Ref.~\cite{eleven}, where the backgrounds were found
to exceed the signal. In contrast, the size of the Higgs 
cross section is smaller in our case, because of the kinematic suppression
 induced by  requiring the detection of the fourth $b-$quark. 
However, 
with an annual luminosity of 100 fb$^{-1}$, expected from
the high luminosity option of the LHC, and a  very good 
 $b$-tagging efficiency, one would obtain a clearly viable signal. 
To illustrate this we show the signal rates on the right-hand 
scale of Fig.~\ref{fig:reso},
for an optimistic $b$-tagging  factor of $\epsilon_b^4=0.1$, 
corresponding to $\epsilon_b=56\%$. Such a high value 
is now considered realistic for the TeV-2 run at FNAL
and can be achieved
by combining the  silicon vertex and the lepton tagging efficiencies
\cite{lastbtag}.
It is also close to the optimistic expectation of a $b$-tagging
efficiency of about 50\% even for the high luminosity run of the LHC 
\cite{notes}.

For the above efficiency and luminosity,
one would obtain between ten and hundred  Higgs events 
per year with signal-to-background ratios  
above one for $M_H^\pm$ as large as $800$~GeV (for $\tan\beta=40$), as shown in
Tab.~\ref{tab:discovery}. This shows the predicted number of events
in a window of 80~GeV centered around the Higgs
resonances for both signal and background, 
together with the corresponding statistical
factors $S/\sqrt B$. By looking at those rates, one would expect 
an accessible signal for $M_{H^\pm}\lsim~600$ GeV in the high $\tan\beta$
region ($\gsim40$). Similar results also hold for the case 
of low $\tan\beta$ values ($\lsim1.5$).

However, given the not too large rates of the surviving Higgs events, 
the actual size of the MSSM parameter space that can be covered
strongly depends on the $b$-tagging efficiency,
$\epsilon_b$. For instance, changing it from 56\% to 40\% would result
in a reduction of $\epsilon_b^4$ by a factor of four.
This corresponds to a suppression of the $S/\sqrt B$ rates of 
Tab.~\ref{tab:discovery} by a factor of two. 
A similar effect would occur if a 30 GeV
cut in transverse momenta of both missing and observable particles is
enforced, as opposed to the 20 GeV value advocated here. In this
case, the suppression would be somewhat smaller though, about a factor of 3 
on the event rates and 1.7 on the statistical significances.

Before concluding, we would like to come back to and justify
what we have mentioned in the Introduction: that the size of the backgrounds 
(\ref{fiveb})--(\ref{single}) is generally smaller than that
of process (\ref{fivea}) considered so far, for the $b$-tagging efficiencies
assumed in this paper. Rather than re-running
all the simulations for each of these additional final states, 
we have assumed the two (anti)top 
quarks to have already been reconstructed, with similar efficiency in each
case. This way, we can compute the $2\to4$ cross sections 
(\ref{fiveb})--(\ref{single}) and apply the transverse momentum, pseudorapidity
and separation cuts of Section \ref{sec_selection} to the jet-jet system
accompanying the $t\bar t$ pair, alongside those of eq.~(\ref{bbcuts}).
We do so also for the case of process (\ref{fivea}). (For all such computations
we have resorted again to MadGraph \cite{tim}.)
By adopting $\epsilon_b=0.56$ and assuming $\epsilon_{g,~q\ne b}=0.01$ to be
the rejection factor against non-$b-$jets, then the 
rates obtained this way scale as follows:
\begin{equation}\label{scale}
\sigma(gg \rightarrow t\bar t b\bar b)
~:~
\sigma(gb\to t\bar t gb)
~:~
\sigma(gg \rightarrow t\bar tgg~+~t\bar t q\bar q)
~\approx~
3.38~:~0.56~:~0.48. 
\end{equation}
Thus, the singly and doubly $b$-mis-tagged backgrounds are both one
order of magnitude smaller than the pure $4b$-process.
The same applies also to the other channel contributing to a possible
{double} $b$-mis-tagging, i.e., 
\begin{equation}
gq\to t\bar t gq,
\end{equation}
where $q\ne b$, whose production rates are in fact comparable to those of
process (\ref{fivea}) (about 25\% smaller).
Altogether, they would add a $40\%$ or so contribution
to the $t\bar t b\bar b$ background.
Some of these backgrounds are likely to be enhanced
by a larger probability of a $c-$quark jet faking a $b$-tag, which we have
not taken into account.
However, their inclusion would not change our main results, so
that we have left these reactions aside for the time being.

\section{Summary and conclusions}
\label{sec_summary}

The discovery at the LHC of charged scalar particles would definitely confirm 
the existence of new physics beyond the SM.
In this respect, a very special r{\^o}le is played by the 
charged Higgs bosons of 2-Higgs Doublet Models (2HDMs), 
as their production and decay dynamics can entirely be described at
tree level by only two parameters.
However, the feasibility of their detection at the LHC
has always been far from certain if the mass of the new particles 
is much larger than the top mass.
Therefore, a high-on-the-list priority is to devise phenomenological
strategies that would allow one to meet the difficult challenge of their 
detection at the LHC collider. 

We have contributed here to this task by considering the production and 
decay of charged Higgs scalars of the MSSM in the channel $gg\to t\bar b H^-
+ \bar t b H^+\to b\bar b t\bar t\to b\bar b b\bar b W^+W^-$,
in which one $W^\pm$ decays hadronically and the other leptonically.
The major feature of our analysis, as compared to others carried
out in the past, is that all four $b-$quarks present in the final state
are required to be recognised as such. The advantage of this procedure is that
it allows one to exploit the differences existing 
between signal and background in the kinematics of the heavy quark jets. In 
fact, in the dominant background, $gg\to t\bar t b\bar b$, the two $b-$quarks
produced in association with the $t\bar t$ pair are soft, collinear
and at rather low invariant mass. In contrast, in the Higgs signal,
at least one of the two  is expected to be energetic and isolated, 
as long as $M_{H^\pm}$ is significantly
larger than $m_t$. The disadvantage is that the additional 
$b-$quark produced in Higgs events has rather low transverse momentum, 
so that its detection
imposes a sizeable loss of signal.

By exploiting a selection procedure that allows one to reconstruct
both top and antitop masses and after imposing tight constraints 
on  the two $b-$quarks accompanying
the $t\bar t$ pair, we do see Higgs peaks appearing on top of a flat
combinatorial background and also above
the continuum from $t\bar t b\bar b$ events.
Their statistical significance is such that viable signals can be obtained
for charged Higgs boson masses up to 600~GeV or so, when $\tan\beta$
is either below 1.5 or above 40, with a total number of Higgs events
of the order of a several tens every 100 inverse femtobarns of luminosity. 

Such mass coverage is significantly higher than that achieved in previous
analyses based on a triple $b$-tagging. However,
we should stress that these conclusions rely on a high, though not 
unrealistic, $b$-tagging efficiency, of about 50\% per single $b-$jet,
and a  transverse momentum cut
on jets and leptons, of 20 GeV, somewhat lower than the
threshold normally considered.
If such performances can be achieved by the LHC detectors while the 
machine is running at high luminosity, then the `lepton plus $4b$'
channel represents a profitable new means to access such elusive
yet crucial particles over significant portions of the
MSSM parameter space.
Besides, this channel will be very useful for the
measurement of the charged Higgs boson parameters, given the higher purity
of the signal in this case. 
Certainly, our results are sufficiently optimistic to justify a more
detailed detector-level study, incorporating hadronisation of
the final state partons, the effects of jet identification
and reconstruction as well as a full background simulation. 

\section*{Acknowledgements}

D.J.M, S.M. and D.P.R. would like to thank the Centre for Particle Theory
and Grey College at Durham University
for the kind hospitality while part of this
work was carried out.
D.J.M and S.M. are grateful to the UK-PPARC for research grants.
D.P.R. thanks the UK-PPARC and IFCPAR for financial support. 
The authors are indebted to K. Odagiri for numerical comparisons and many
fruitful discussions.

\vfill\clearpage\thispagestyle{empty}

\begin{table}[b]
\begin{center}
\begin{tabular}{|c|c|c|c|c|c|}
\hline
\multicolumn{6}{|c|}
{Signal, $\tan\beta=30$ (fb)}\\
\hline
$M_{A(H^\pm)}$ & 05A & 09A & 10A & 01A & 07A \\
\hline
$200(214)$ & 
$217.14 \pm 0.18$ &
$226.67 \pm 0.20$ &
$199.11 \pm 0.15$ &
$204.48 \pm 0.17$ &
$222.79 \pm 0.18$ \\
$300(310)$ & 
$123.20 \pm 0.10$ &
$130.79 \pm 0.11$ &
$110.90 \pm 0.086$ &
$116.42 \pm 0.10$ &
$125.76 \pm 0.11$ \\
$400(407)$ & 
$60.414 \pm 0.059$ &
$65.256 \pm 0.066$ &
$53.410 \pm 0.049$ &
$57.333 \pm 0.057$ &
$61.360 \pm 0.059$ \\
$500(506)$ & 
$30.802 \pm 0.044$ &
$33.895 \pm 0.047$ &
$26.742 \pm 0.040$ &
$29.481 \pm 0.043$ &
$31.252 \pm 0.045$ \\
$600(605)$ & 
$16.341 \pm 0.029$ &
$18.268 \pm 0.032$ &
$13.977 \pm 0.026$ &
$15.670 \pm 0.029$ &
$16.455 \pm 0.029$ \\
\hline\hline

\multicolumn{6}{|c|}
{Background (fb)} \\
\hline
$$ & 05A & 09A & 10A & 01A & 07A \\
\hline
$$ & 
$1863\pm56$ & 
$1949\pm59$ & 
$1706\pm51$ & 
$1755\pm53$ & 
$1901\pm57$ \\
\hline\hline
\multicolumn{6}{|c|}
{MRS-LO [~$Q=\mu=\sqrt{\hat s}$~]} \\
\hline\hline
\multicolumn{6}{|c|}
{$4b~+~2~{\mathrm{jets}}~+~\ell^\pm~+~p^T_{\mathrm{miss}}$
\qquad \qquad 
\qquad \qquad 
\qquad \qquad 
\qquad \qquad 
\qquad \qquad 
No cuts}
\\ \hline
\end{tabular}
\end{center}
\caption{Production cross section for the signal (\ref{four}) (and its
charge conjugate) in the  decay channel (\ref{signature})
(top), for $\tan\beta=30$ and five values of $M_{A(H^\pm)}$ 
 (given in GeV) in the heavy mass range, 
as obtained by using five different sets of PDFs.
Corresponding rates for the background (\ref{fivea}) yielding the same
signature (\ref{signature}) are also given. No cuts have been
implemented. The renormalisation and factorisation scales 
are set equal to the partonic CM energy. Errors are as given by 
{\tt VEGAS}.}
\label{tab:PDF}
\end{table}

\vfill\clearpage\thispagestyle{empty}

\begin{table}[b]
\begin{center}
\begin{tabular}{|c||c|c|c|}
\hline
\multicolumn{4}{|c|}
{Number of events per year}\\
\hline
$M_{H^\pm}\pm40$ GeV & $S$ & $B$ & $S/\sqrt B$ \\
\hline     
$310$ &      
$127.80$ & 
$105.14$ & 
$12.46$ \\
$~~~$ &
$57.80$ & 
$41.90$ & 
$8.92$ \\
$407$ & 
$88.56$ & 
$67.63$ & 
$10.76$ \\
$~~~$ &
$53.78$ & 
$39.21$ & 
$8.58$ \\
$506$ & 
$51.46$ & 
$38.74$ & 
$8.26$ \\
$~~~$ &      
$36.32$ & 
$26.86$ & 
$7.00$ \\
$605$ &  
$29.43$ & 
$21.49$ & 
$6.34$ \\
$~~~$ &
$22.70$ & 
$16.58$ & 
$5.57$ \\
$704$ & 
$17.09$ &      
$11.98$ &      
$4.93$ \\    
$~~~$ &
$14.02$ & 
$9.89$ &     
$4.45$ \\    
$803$ & 
$10.03$ & 
$6.75$ &   
$3.86$ \\
$~~~$ &
$8.61$ & 
$5.81$ & 
$3.57$ \\
\hline\hline
\multicolumn{4}{|c|}
{MRS-LO [~$Q=\mu=\sqrt{\hat s}$~]} \\
\hline\hline
\multicolumn{4}{|c|}
{$4b~+~2~{\mathrm{jets}}~+~\ell^\pm~+~p^T_{\mathrm{miss}}$
\qquad \qquad 
After all cuts}
\\ \hline
\end{tabular}
\end{center}
\caption{Number of events from  
signal (\ref{four}) (and its
charge conjugate) ($S$) and background (\ref{fivea}) ($B$)
in the decay channel (\ref{signature}), along with the
statistical significance (${\protect{S/\sqrt B}}$), 
per 100 inverse femtobarns of integrated luminosity, in a window of
80~GeV around a few selected values of
 $M_{H^\pm}$ (given in GeV) in the heavy mass range, 
with $\tan\beta=40$. Four $b-$jets are assumed to be tagged
with overall efficiency $\epsilon_b^4=0.1$, i.e., $\epsilon_b=56\%$.
All cuts  
discussed in the text, (\ref{pTj})--(\ref{R}),  (a)--(d) and (\ref{bbcuts}),
have been implemented. The renormalisation and factorisation scales 
are set equal to the partonic CM energy. 
The first row corresponds to the $M_4(H)$ distribution whereas
the second refers to the $M_2(H)$ one (see Fig.~\ref{fig:reso}).}
\label{tab:discovery}
\end{table}

\vfill\clearpage\thispagestyle{empty}

\vfill\clearpage\thispagestyle{empty}

\vfill\clearpage\thispagestyle{empty}

\vfill\clearpage\thispagestyle{empty}

\end{document}